\documentclass[onecolumn,unpublished]{quantumarticle}
\usepackage{amsmath}
\usepackage{amssymb}
\usepackage{amsthm}
\usepackage{amsfonts}
\usepackage[caption=false]{subfig}
\usepackage[colorlinks]{hyperref}
\usepackage[all]{hypcap}
\usepackage{tikz}
\usepackage{color,soul}
\usepackage[utf8]{inputenc}
\usepackage{capt-of}
\usepackage[numbers]{natbib}

\newcommand{\hide}[1]{}

\DeclareFixedFont{\ttb}{T1}{txtt}{bx}{n}{12} 
\DeclareFixedFont{\ttm}{T1}{txtt}{m}{n}{12}  

\usepackage{listings}

\renewcommand\thesection{\arabic{section}}

\newcommand{\eq}[1]{\hyperref[eq:#1]{Equation~\ref*{eq:#1}}}
\renewcommand{\sec}[1]{\hyperref[sec:#1]{Section~\ref*{sec:#1}}}
\DeclareRobustCommand{\app}[1]{\hyperref[app:#1]{Appendix~\ref*{app:#1}}}
\newcommand{\fig}[1]{\hyperref[fig:#1]{Figure~\ref*{fig:#1}}}
\newcommand{\tbl}[1]{\hyperref[tbl:#1]{Table~\ref*{tbl:#1}}}
\newcommand{\theoremref}[1]{\hyperref[theorem:#1]{Theorem~\ref*{theorem:#1}}}
\newcommand{\definitionref}[1]{\hyperref[definition:#1]{Theorem~\ref*{definition:#1}}}

\newcommand{\distone}{{d_1}}

\input{qcircuit}

\begin{document}
\title{Flexible layout of surface code computations using AutoCCZ states}

\date{\today}
\author{Craig Gidney}
\email{craiggidney@google.com}
\affiliation{Google Inc., Santa Barbara, California 93117, USA}
\author{Austin G. Fowler}
\affiliation{Google Inc., Santa Barbara, California 93117, USA}

\begin{abstract}
We construct a self-correcting CCZ state (the ``AutoCCZ") with embedded delayed choice CZs for completing gate teleportations.
Using the AutoCCZ state we create efficient surface code spacetime layouts for both a depth-limited circuit (a ripply-carry addition) and a Clifford-limited circuit (a QROM read).
Our layouts account for distillation and routing, are based on plausible physical assumptions for a large-scale superconducting qubit platform, and suggest that circuit-level Toffoli parallelism (e.g. using a carry-lookahead adder instead of a ripple-carry adder) will not reduce the execution time of computations involving fewer than five million physical qubits.
We reduce the spacetime volume of delayed choice CZs by a factor of 4 compared to techniques from previous work (Fowler 2012), and make several improvements to the CCZ magic state factory from (Gidney 2019).
\end{abstract}

\maketitle

\section{Introduction}
\label{sec:introduction}

An interesting consequence of the topological nature of quantum computation in the surface code \cite{fowler2012surfacecodereview} is that, in spacetime diagrams describing the computation, there is very little distinction between spacelike and timelike directions.
It is valid to redirect a qubit's worldline ``backwards in time" so that its next operation actually happened earlier.
There is of course a forward-in-time description of the computation occurring, where a magic state representing an operation is prepared, applied to the qubit via gate teleportation, and then completed via classically controlled fixup operations.
But conceptually it is helpful to treat the time direction as just another space direction.

Computations that only involve Clifford operations are particularly easy to treat purely topologically, with time equivalent to space, because when Clifford operations are applied via gate teleportation the resulting fixup operations are always classically controlled Paulis (which can be applied entirely within the classical control system).
In order to treat time like space when performing more general operations, such as T gates and Toffoli gates, it is necessary to use techniques such as the selective routing construction described by Fowler in \cite{fowler2012time}.

Fowler's technique allows general operations to be laid out arbitrarily through spacetime, but produces a series of ``routing qubits" that must be stored until an adaptive measurement process determines whether to measure them in the X or Z basis.
The measurement depth (e.g. T depth) of the circuit determines how many times the classical control system will have to perform a set of measurements, decide which basis to use for the next set of measurements, and start those measurements.
The speed at which the control system can run this loop, and work through the measurements, determines the speed of computation.

The characteristic time it takes the control system to react to a measurement, and perform the following dependent measurement, is the control system's ``reaction time".
We refer to the general paradigm of performing a quantum computation whose speed is limited only by the measurement depth of the circuit and the reaction time of the control system as ``reaction limited computation".

\begin{figure}
    \resizebox{\linewidth}{!}{
        \includegraphics{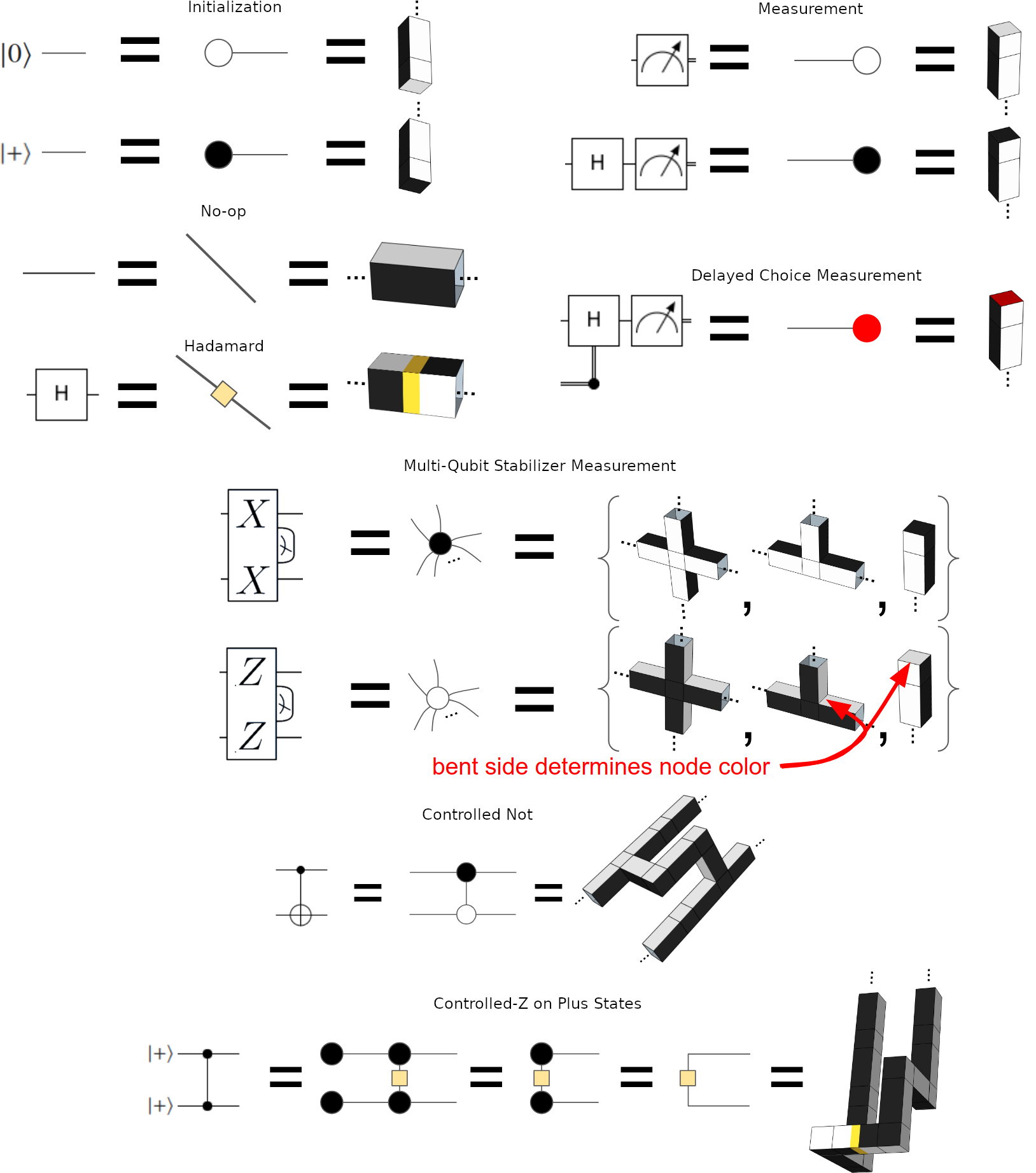}
    }
    \caption{
        \label{fig:circuit-zx-lattice-rosetta}
        Equivalent concepts expressed in quantum circuit diagrams, ZX calculus graphs, and 3d topological diagrams.
        In circuit diagrams, time goes from left to right.
        In ZX calculus graphs, there is no preferred time direction.
        In 3d topological diagrams, times goes from bottom to top.
        We never show Pauli operations in lattice surgery diagrams or in ZX calculus graphs, because they are performed by the classical control system instead of by operating on qubits.
        Our usage of the ZX calculus is somewhat non-standard in that we consider ZX graphs to be equivalent if they are equal modulo Pauli operations, we use a non-standard node coloring that matches the coloring of our topological diagrams, and we introduce a ``delayed choice node" to represent adaptive effects coming from the classical control system.
        We exaggerate the spacing of our 3d topological diagrams, as in \cite{gidney2018magic}, so that it is possible to see how the components are interconnected.
    }
\end{figure}

Our contributions in this paper relate to decreasing the space overhead of reaction limited computation, and making it easier to route.
The paper is organized as follows.
In \sec{introduction} we describe the context in which we are operating and note some notational conventions.
In \sec{delayed-cz} we present our optimized version of a reaction limited selective CZ, which we refer to as a delayed choice CZ.
In \sec{auto-ccz} we show how to produce and consume AutoCCZ states, which make routing easier because they decouple the consumption of the CCZ state from the fixup operations needed to complete a gate teleportation.
\sec{ccz-factory} presents several improvements to the CCZ distillation factory from \cite{gidney2018magic}.
In \sec{adder} we lay out a reaction limited ripple-carry adder that relies heavily on AutoCCZ states.
In \sec{lookup} we lay out a QROM read that uses multiple access hallways in order to \emph{nearly} transition from being Clifford limited to being reaction limited.
Finally, in \sec{conclusion} we summarize our contributions and discuss their implications.

In this paper we will represent quantum computations using quantum circuit diagrams, ZX calculus graphs \cite{de2017zxlattice} and 3d topological diagrams.
\fig{circuit-zx-lattice-rosetta} shows how to translate between the three representations.

When making estimates, we will be using assumptions that are plausible for a future large-scale superconducting qubit platform: a reaction time of 10 microseconds, a surface code cycle time of 1 microsecond, and a characteristic gate error rate of $10^{-3}$.
We show these quantities in \tbl{assumptions}, which also notes the effect of changing each value.

\begin{table}
\resizebox{\linewidth}{!}{
  \begin{tabular}{r|c|l|l|}
    Quantity & Value & Effect of 10x decrease & Effect of 10x increase \\
\hline\hline
   Physical gate error rate
        &$10^{-3}$
        &4x less space, same time, fewer factories
        &Too close to threshold for tractable computation
        \\
        &
        &Reaction limited at 7 factories
        &
        \\\hline
  Surface code cycle time
        &1 $\mu$s
        &Same time, fewer factories
        &10x more spacetime volume
        \\
        &
        &Reaction limited at 2 factories
        &Reaction limited at 135 factories
        \\\hline
  Reaction time
        &10 $\mu$s
        &Easy to trade space for time
        &Hard to trade space for time
        \\
        &
        &Reaction limited at 135 factories
        &Reaction limited at 2 factories
        \\\hline
  Physical connectivity
        &planar
        &N/A
        &N/A
        \\\hline
  \end{tabular}
}
  \caption{
    \label{tbl:assumptions}
    Physical assumptions we make in this paper, and the effect of varying them.
  }
\end{table}

\section{A compact delayed choice CZ}
\label{sec:delayed-cz}

\begin{figure}[h!]
    \resizebox{\linewidth}{!}{
        \includegraphics{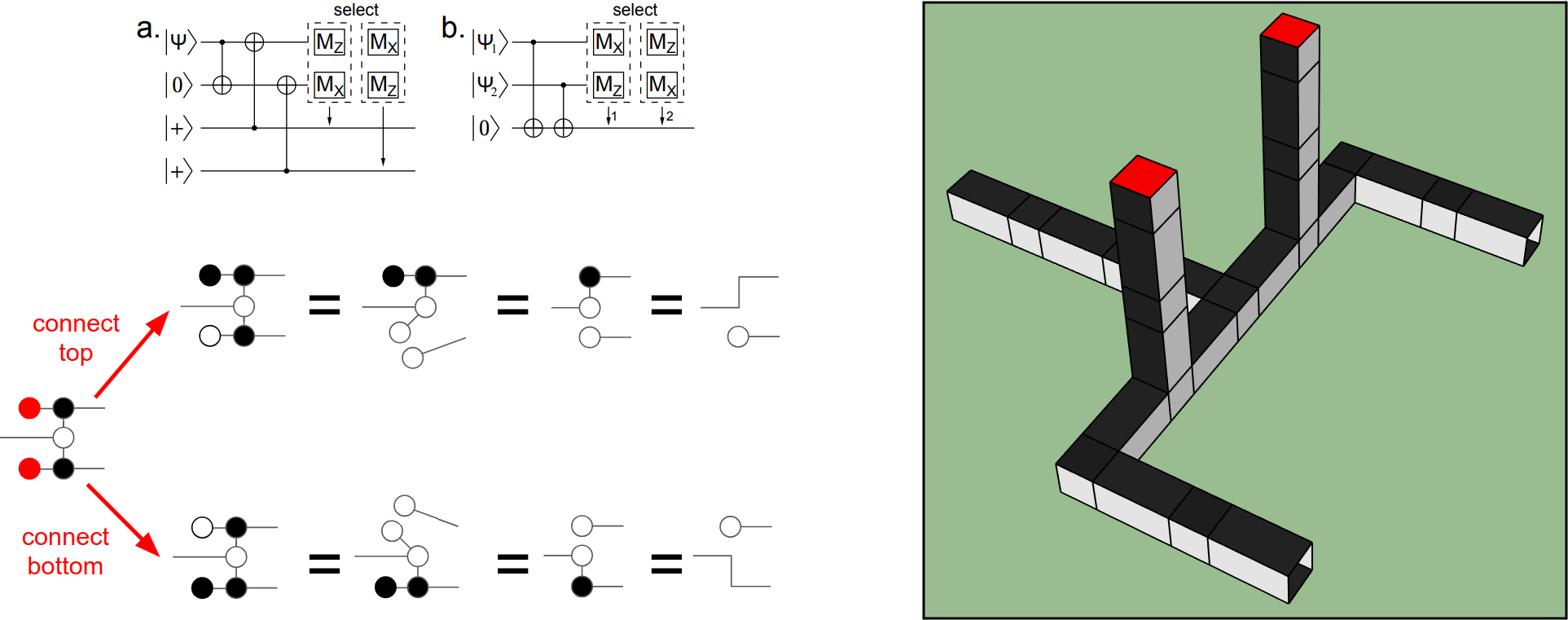}
    }
    \caption{
        \label{fig:precompute-switch}
        Delayed choice multiplex/demultiplex construction from \cite{fowler2012time}.
        Top left is a circuit diagram directly from \cite{fowler2012time}.
        The bottom left shows the process as a ZX calculus graph which, unlike the circuit diagram, is identical for multiplexing and demultiplexing.
        Known rewrite rules are used to show equivalence with the claimed ``choose which route is connected" functionality.
        On the right side is a 3d topological diagram of a lattice surgery implementation of the construction.
        The vertical poles coming out of the branches of the fork are the routing qubits used to control which of the two branches connects to the rear trunk.
        They can be extended arbitrarily.
        The red squares atop the routing qubit columns are placeholders for an eventual X or Z basis measurement, to be determined by classical control software.
    }
\end{figure}

Fowler's selective routing technique from \cite{fowler2012time} is based on controllable multiplexers and demultiplexers (see \fig{precompute-switch}).
Each (de)multiplexer produces two ``routing qubits" which, depending on whether they are measured in the X or Z basis at a later time, can route a data qubit through one of multiple precomputed worldlines.
By chaining these selective computations together through space instead of through time, the computation becomes reaction limited.

For example, consider a series of Toffoli gates where the output of each gate affects the control of a following Toffoli gate, such as the Toffolis in a ripple-carry adder.
Normally, the current Toffoli would need to be completely finished (including fixup operations necessitated due to using gate teleportation) before it was possible to start the next Toffoli gate.
But, by using multiplexers routing through multiple possible precomputed fixups, it is possible to apply all of the Toffoli operations simultaneously while delayed-choice routing through all the various possible fixups.

\begin{figure}[h!]
    \resizebox{\linewidth}{!}{
        \includegraphics{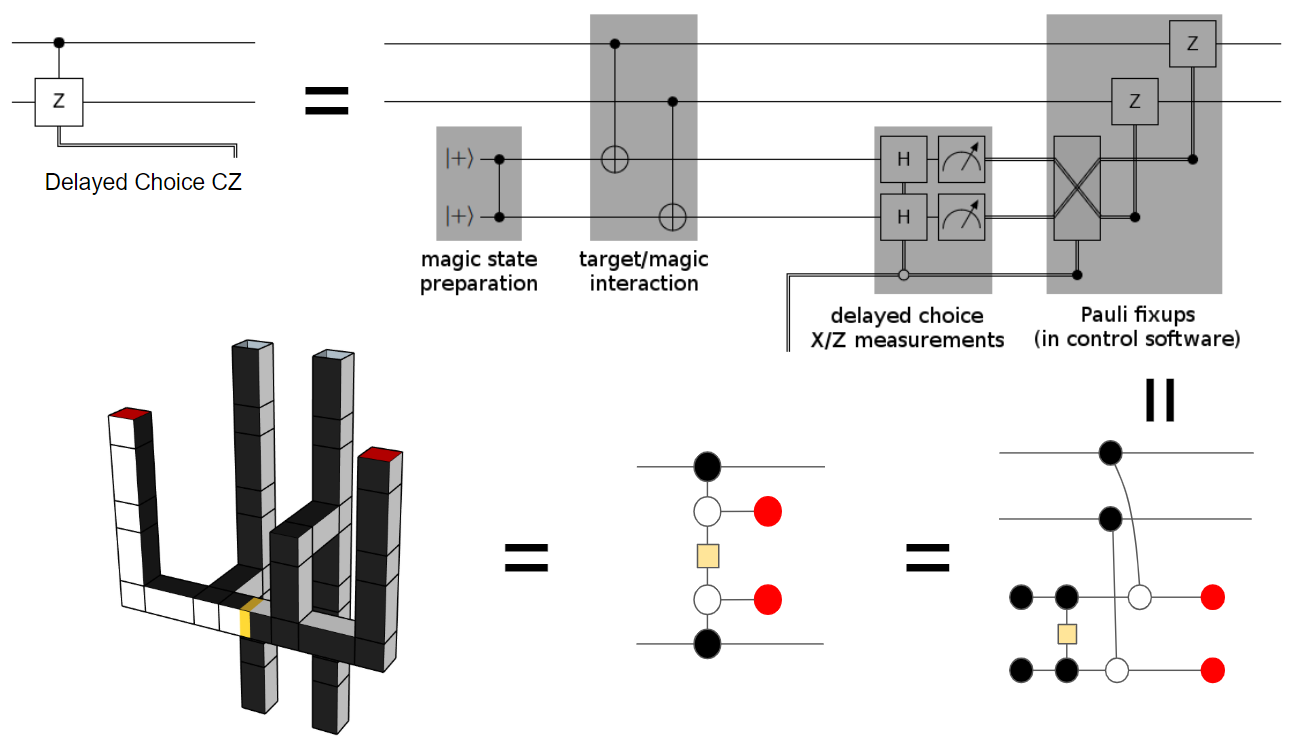}
    }
    \caption{
        \label{fig:delayed-choice-cz}
        Our optimized delayed choice CZ as a circuit and as a lattice surgery construction.
        The two forms are shown to be equivalent via ZX graph identities.
        During execution, the choice of whether or not to apply the CZ is delayed by extending the red-topped columns (the ``routing qubits") in the 3d topological diagram.
        Once the choice is known, the columns are terminated with the red square replaced either by a white square (activates the CZ) or a black square (skips the CZ).
        The circuit can be opened in the online simulator Quirk by \href{https://algassert.com/quirk\#circuit=\%7B\%22cols\%22\%3A\%5B\%5B1\%2C1\%2C\%22\%E2\%80\%A2\%22\%2C\%22Z\%22\%5D\%2C\%5B\%22Amps2\%22\%5D\%2C\%5B\%5D\%2C\%5B\%22\%E2\%80\%A2\%22\%2C1\%2C\%22X\%22\%5D\%2C\%5B1\%2C\%22\%E2\%80\%A2\%22\%2C1\%2C\%22X\%22\%5D\%2C\%5B1\%2C1\%2C1\%2C1\%2C\%22Measure\%22\%5D\%2C\%5B1\%2C1\%2C\%22H\%22\%2C\%22H\%22\%2C\%22\%E2\%97\%A6\%22\%5D\%2C\%5B1\%2C1\%2C\%22Measure\%22\%2C\%22Measure\%22\%5D\%2C\%5B1\%2C1\%2C\%22\%3C\%3C2\%22\%2C1\%2C\%22\%E2\%80\%A2\%22\%5D\%2C\%5B1\%2C\%22Z\%22\%2C1\%2C\%22\%E2\%80\%A2\%22\%5D\%2C\%5B\%22Z\%22\%2C1\%2C\%22\%E2\%80\%A2\%22\%5D\%2C\%5B\%22Amps2\%22\%2C1\%2C1\%2C1\%2C\%22\%E2\%97\%A6\%22\%5D\%2C\%5B\%5D\%2C\%5B\%22Amps2\%22\%2C1\%2C1\%2C1\%2C\%22\%E2\%80\%A2\%22\%5D\%5D\%2C\%22init\%22\%3A\%5B\%22\%2B\%22\%2C\%22\%2B\%22\%2C\%22\%2B\%22\%2C\%22\%2B\%22\%2C\%22\%2B\%22\%5D\%7D}{following this link}.
    }
\end{figure}

Although the (de)multiplexer construction from \cite{fowler2012time} is very general, it is often not optimal.
For example, teleporting a CCZ gate produces up to three possible CZ fixup operations.
Using the multiplexer construction to delay the choice of whether or not the various CZ fixups should be applied would produce eight routing qubits per potential CZ (because there are two qubits involved in a CZ and each must go through a multiplexer/demultiplexer pair).
In \fig{delayed-choice-cz}, we present a more efficient construction for performing delayed choice CZs that only uses two routing qubits.

\section{The AutoCCZ state}
\label{sec:auto-ccz}

We can embed three instances of the delayed choice CZ construction directly into a CCZ state, so that there is one delayed choice CZ for each CZ fixup that may be needed when performing gate teleportation.
This augments the CCZ state into an ``AutoCCZ" state, so named because it automatically cleans up its own CZ fixup garbage.
This makes consuming the state much simpler, because no corrections are needed at the consumption site.
(Note that any magic state can be augmented into an auto magic state.
For example, Figure 17 of \cite{litinski2018} defines an auto-corrected T gate.)

\begin{figure}[h!]
    \resizebox{\linewidth}{!}{
        \includegraphics{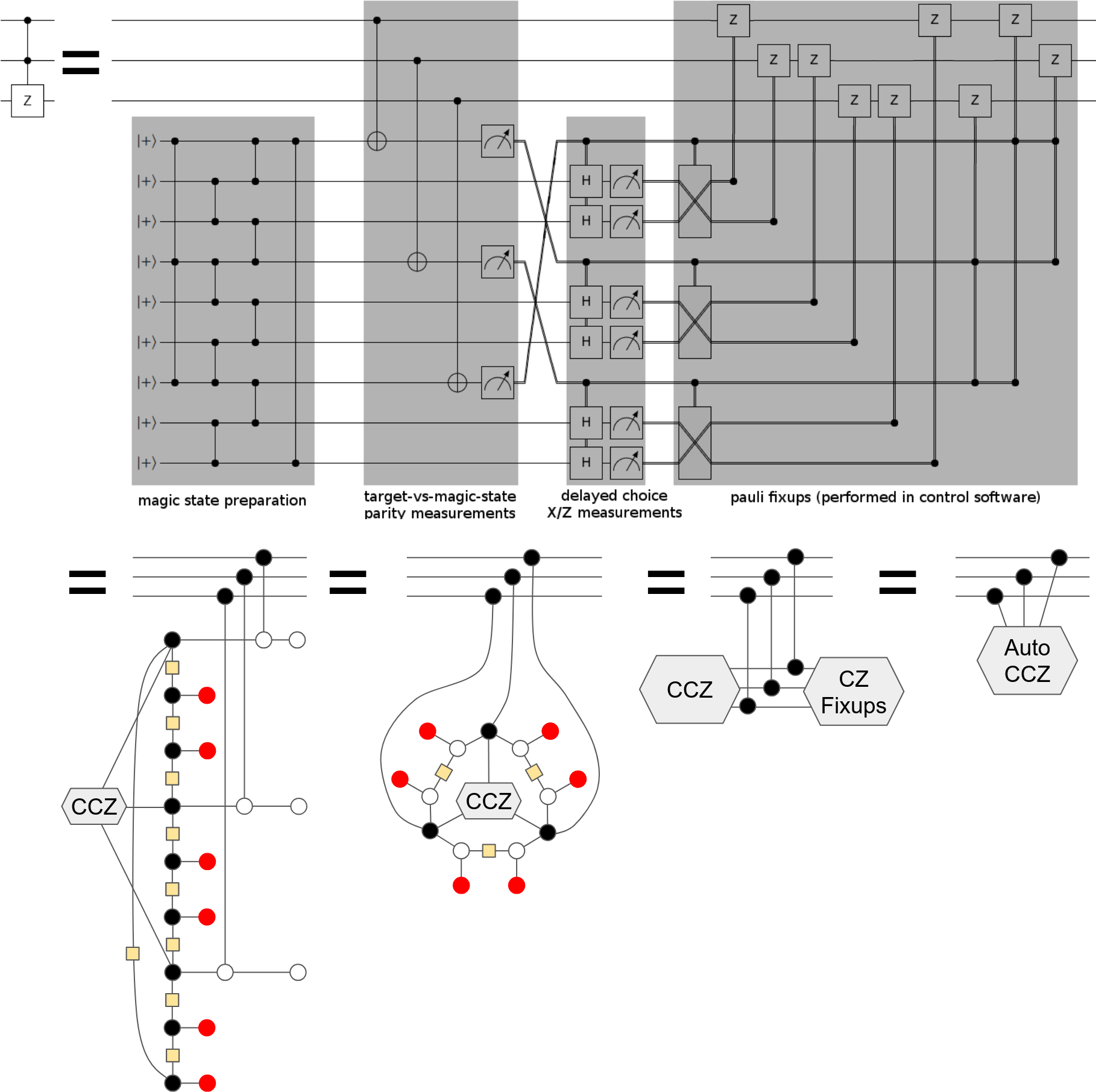}
    }
    \caption{
        \label{fig:auto-ccz-circuit}
        A circuit diagram and ZX calculus graph simplification showing how to create and consume an AutoCCZ magic state, powered by delayed choice CZs, to perform a CCZ gate.
        The bottom left part of the circuit is producing the AutoCCZ magic state, then the middle left does parity measurements vs the target qubits, then the middle right uses those measurements to determine the basis of measurements that determine whether fixup operations occur, then finally on the right side all the measurement results are used to update the Pauli frame tracked in the control software.
        The circuit can be opened in the online simulator Quirk by \href{https://algassert.com/quirk\#circuit=\%7B\%22cols\%22\%3A\%5B\%5B1\%2C1\%2C1\%2C\%22\%E2\%80\%A2\%22\%2C1\%2C1\%2C\%22\%E2\%80\%A2\%22\%2C1\%2C1\%2C\%22Z\%22\%5D\%2C\%5B1\%2C1\%2C1\%2C1\%2C\%22~cz\%22\%2C1\%2C\%22~cz\%22\%2C1\%2C\%22~cz\%22\%2C1\%2C\%22~cz\%22\%5D\%2C\%5B1\%2C1\%2C1\%2C\%22~cz\%22\%2C1\%2C\%22~cz\%22\%2C1\%2C\%22~cz\%22\%2C1\%2C\%22~cz\%22\%5D\%2C\%5B1\%2C1\%2C1\%2C\%22\%E2\%80\%A2\%22\%2C1\%2C1\%2C1\%2C1\%2C1\%2C1\%2C1\%2C\%22Z\%22\%5D\%2C\%5B\%22Amps3\%22\%5D\%2C\%5B\%5D\%2C\%5B\%22\%E2\%80\%A2\%22\%2C1\%2C1\%2C\%22X\%22\%5D\%2C\%5B1\%2C\%22\%E2\%80\%A2\%22\%2C1\%2C1\%2C1\%2C1\%2C\%22X\%22\%5D\%2C\%5B1\%2C1\%2C\%22\%E2\%80\%A2\%22\%2C1\%2C1\%2C1\%2C1\%2C1\%2C1\%2C\%22X\%22\%5D\%2C\%5B1\%2C1\%2C1\%2C\%22Measure\%22\%2C1\%2C1\%2C\%22Measure\%22\%2C1\%2C1\%2C\%22Measure\%22\%5D\%2C\%5B1\%2C1\%2C1\%2C1\%2C1\%2C1\%2C\%22Swap\%22\%2C1\%2C1\%2C\%22Swap\%22\%5D\%2C\%5B1\%2C1\%2C1\%2C\%22Swap\%22\%2C1\%2C1\%2C\%22Swap\%22\%5D\%2C\%5B1\%2C1\%2C1\%2C\%22\%E2\%80\%A2\%22\%2C\%22H\%22\%2C\%22H\%22\%5D\%2C\%5B1\%2C1\%2C1\%2C1\%2C1\%2C1\%2C\%22\%E2\%80\%A2\%22\%2C\%22H\%22\%2C\%22H\%22\%5D\%2C\%5B1\%2C1\%2C1\%2C1\%2C1\%2C1\%2C1\%2C1\%2C1\%2C\%22\%E2\%80\%A2\%22\%2C\%22H\%22\%2C\%22H\%22\%5D\%2C\%5B1\%2C1\%2C1\%2C1\%2C\%22Measure\%22\%2C\%22Measure\%22\%2C1\%2C\%22Measure\%22\%2C\%22Measure\%22\%2C1\%2C\%22Measure\%22\%2C\%22Measure\%22\%5D\%2C\%5B1\%2C1\%2C1\%2C\%22\%E2\%80\%A2\%22\%2C\%22\%3C\%3C2\%22\%5D\%2C\%5B1\%2C1\%2C1\%2C1\%2C1\%2C1\%2C\%22\%E2\%80\%A2\%22\%2C\%22\%3C\%3C2\%22\%5D\%2C\%5B1\%2C1\%2C1\%2C1\%2C1\%2C1\%2C1\%2C1\%2C1\%2C\%22\%E2\%80\%A2\%22\%2C\%22\%3C\%3C2\%22\%5D\%2C\%5B\%22Z\%22\%2C1\%2C1\%2C1\%2C\%22\%E2\%80\%A2\%22\%5D\%2C\%5B1\%2C\%22Z\%22\%2C1\%2C1\%2C1\%2C\%22\%E2\%80\%A2\%22\%5D\%2C\%5B1\%2C\%22Z\%22\%2C1\%2C1\%2C1\%2C1\%2C1\%2C\%22\%E2\%80\%A2\%22\%5D\%2C\%5B1\%2C1\%2C\%22Z\%22\%2C1\%2C1\%2C1\%2C1\%2C1\%2C\%22\%E2\%80\%A2\%22\%5D\%2C\%5B1\%2C1\%2C\%22Z\%22\%2C1\%2C1\%2C1\%2C1\%2C1\%2C1\%2C1\%2C\%22\%E2\%80\%A2\%22\%5D\%2C\%5B\%22Z\%22\%2C1\%2C1\%2C1\%2C1\%2C1\%2C1\%2C1\%2C1\%2C1\%2C1\%2C\%22\%E2\%80\%A2\%22\%5D\%2C\%5B1\%2C1\%2C\%22Z\%22\%2C1\%2C1\%2C1\%2C\%22\%E2\%80\%A2\%22\%2C1\%2C1\%2C\%22\%E2\%80\%A2\%22\%5D\%2C\%5B\%22Z\%22\%2C1\%2C1\%2C\%22\%E2\%80\%A2\%22\%2C1\%2C1\%2C1\%2C1\%2C1\%2C\%22\%E2\%80\%A2\%22\%5D\%2C\%5B1\%2C\%22Z\%22\%2C1\%2C\%22\%E2\%80\%A2\%22\%2C1\%2C1\%2C\%22\%E2\%80\%A2\%22\%5D\%2C\%5B\%22Amps3\%22\%5D\%5D\%2C\%22gates\%22\%3A\%5B\%7B\%22id\%22\%3A\%22~cz\%22\%2C\%22name\%22\%3A\%22CZ\%22\%2C\%22circuit\%22\%3A\%7B\%22cols\%22\%3A\%5B\%5B\%22\%E2\%80\%A2\%22\%2C\%22Z\%22\%5D\%5D\%7D\%7D\%5D\%2C\%22init\%22\%3A\%5B\%22\%2B\%22\%2C\%22\%2B\%22\%2C\%22\%2B\%22\%2C\%22\%2B\%22\%2C\%22\%2B\%22\%2C\%22\%2B\%22\%2C\%22\%2B\%22\%2C\%22\%2B\%22\%2C\%22\%2B\%22\%2C\%22\%2B\%22\%2C\%22\%2B\%22\%2C\%22\%2B\%22\%5D\%7D}{following this link}.
    }
\end{figure}

\begin{figure}[h]
    \resizebox{\linewidth}{!}{
        \includegraphics{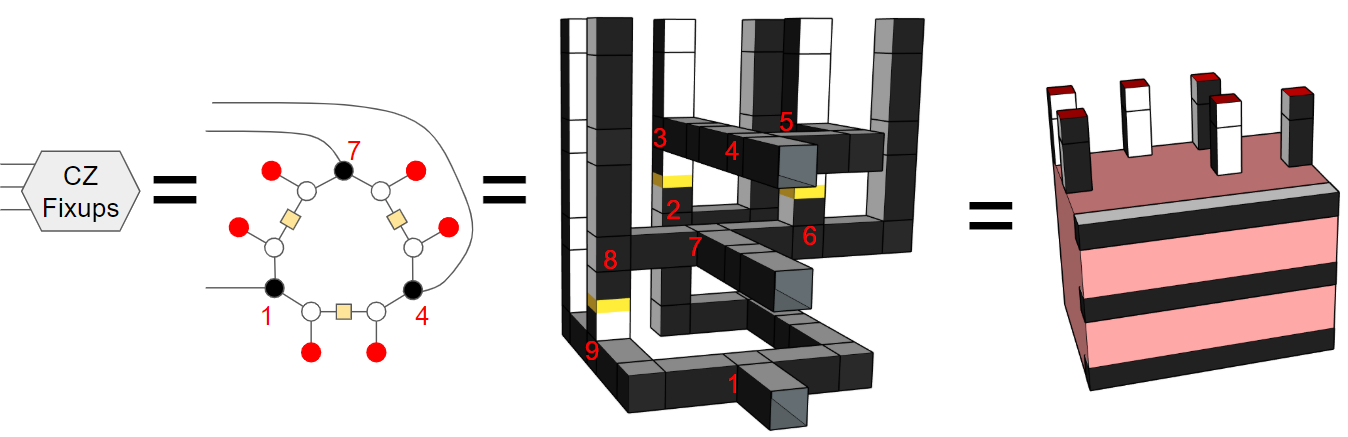}
    }
    \caption{
        \label{fig:auto-ccz-fixup-lattice}
        Lattice surgery implementation of the CZ fixups bubble from \fig{auto-ccz-circuit}.
        The 3d model is stored in ancillary file ``ccz-fixup.skp"; it can be opened online using Sketchup.
        The tee-junctions in the 3d diagram have been numbered, so that it is easier to see how they correspond to the cycle in the ZX calculus graph.
        The pink box on the right, with routing qubit ``chimneys", is the simplified abstract representation that we will use in larger diagrams.
    }
\end{figure}

In \fig{auto-ccz-circuit}, we show a circuit diagram producing and consuming an AutoCCZ state.
The figure also shows how to represent this concept as a ZX calculus graph.
Then, in \fig{auto-ccz-fixup-lattice}, we construct a compact ``CCZ fixup box" with contents equivalent to the CZ fixup part of the ZX graph.
AutoCCZ states are produced by linking the CCZ state output from a CCZ factory to the ports of the fixup box as the CCZ state is routed to its destination.

\begin{figure}[h]
    \resizebox{\linewidth}{!}{
        \includegraphics{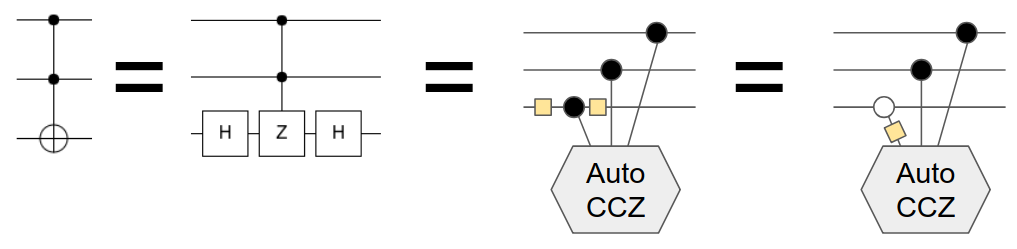}
    }
    \caption{
        \label{fig:toffoli-from-ccz}
        Performing a Toffoli gate by consuming an AutoCCZ state.
    }
\end{figure}

\begin{figure}[h]
    \resizebox{\linewidth}{!}{
        \includegraphics{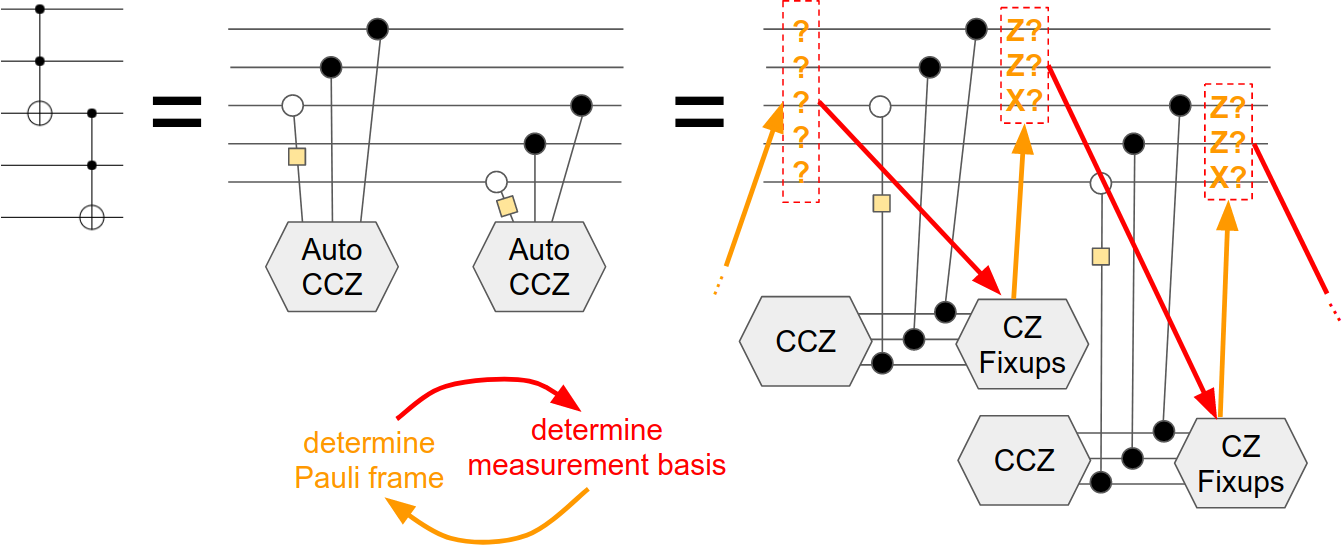}
    }
    \caption{
        \label{fig:toffoli-sweep}
        Performing a series of Toffoli gates using reaction limited quantum computation \cite{fowler2012time}.
        The Toffolis can be laid out in a spacelike fashion, so that they are ``performed simultaneously", but doing so will produce routing qubits that the control software will still process serially.
        The rate at which the control software can cycle between computing Pauli frames and applying fixups, i.e. the time it takes to measure a set of routing qubits and determine the basis to measure the next set in, determines the speed of the quantum computation.
    }
\end{figure}

In \fig{toffoli-from-ccz}, we show how to use a CCZ state to perform a Toffoli operation.
In \fig{toffoli-sweep}, we show the resulting reaction limited control system cycle.

We believe that it is possible to reduce the amount of spacetime volume we use to transform CCZ states into AutoCCZ states.
For example, our current construction uses six routing qubits (two for each possible CZ fixup) but intuition would suggest that only three qubits should be needed in order to distinguish between the eight possible fixup cases.
Alternatively, it should be possible to save volume by carefully integrating the delayed choice CZ fixups directly into the CCZ factory.
We leave this task as future work.

\section{Improvements to CCZ distillation}
\label{sec:ccz-factory}

Operations that are not native to the surface code can be performed using magic state distillation \cite{bravyi2005distillation} and gate teleportation \cite{gottesman1999gateteleport}.
A particularly useful magic state is the CCZ state $|CCZ\rangle = CCZ |+\rangle^{\otimes 3} = \sum_{a,b,c \in \{0,1\}} (-1)^{abc} |abc\rangle$.
The reason this state is useful is because the quantum equivalent of the AND gate, the Toffoli gate, is not native to the surface code but can be performed by consuming one CCZ state \cite{jones2013, eastin2013distilling, gidney2018magic} (see \fig{toffoli-from-ccz}).
Algorithms with a lot of arithmetic, such as Grover's algorithm and Shor's algorithm, perform many Toffoli gates and benefit from using a state specialized to this task.

It is also possible to perform Toffoli gates using T states, but four states are required instead of one \cite{jones2013}.
Whether this is better or worse than using a CCZ state, or some other technique, depends on the relative spacetime volumes of the different types of magic state factories, their error rates, and the number of operations that need to be performed.

In this paper we will be using the CCZ factory from \cite{gidney2018magic}.
It produces level 0 T states using the state injection technique of Li \cite{li2015}, then distills level 1 T states using one round of 15-to-1 distillation based on the Reed-Muller code \cite{bravyi2005distillation}, then uses those T states to perform an error-detecting Toffoli \cite{jones2013} to produce the final CCZ states.

We make several small improvements to this factory, in order to improve its depth:

\begin{itemize}
    \item We choose code distances based on the target error rate, instead of based on what packs most neatly.
    \item We use six level 1 T factories instead of five.
    \item We improve the depth of the T teleportation/injection construction.
    \item We interleave the level 1 T factories better.
\end{itemize}

\begin{figure}[h]
    \resizebox{0.8 \linewidth}{!}{
        \includegraphics{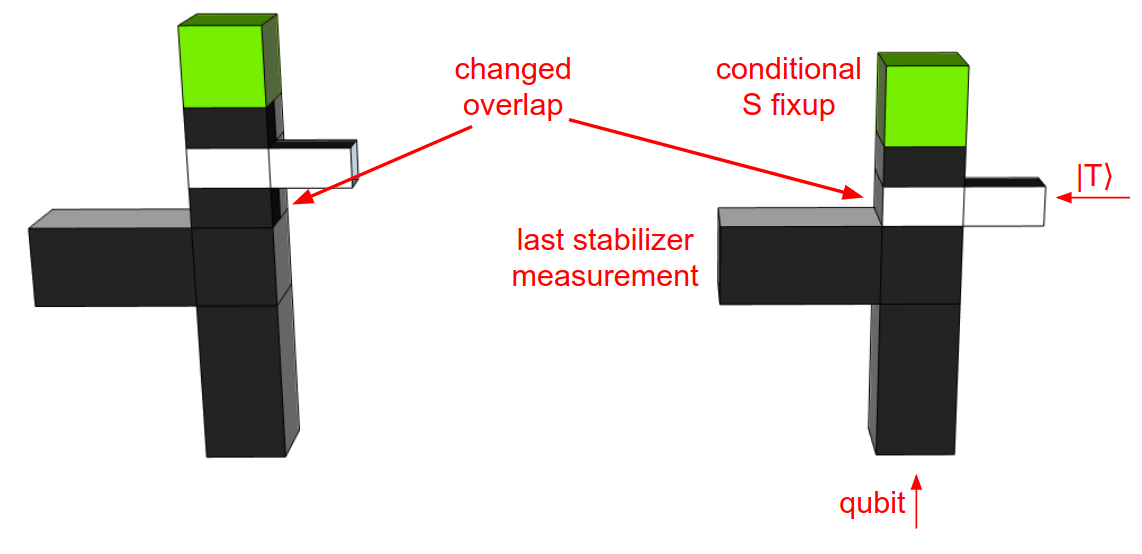}
    }
    \caption{
        \label{fig:t-inject-overlap}
        Improved method for gate teleporting a T state into a qubit.
        This 3d model is stored in ancillary file ``inject-t.skp"; it can be opened online using Sketchup.
        The new method (right) saves $0.5d$ surface code cycles, where $d$ is the code distance of the factory, compared to the old method (left) \cite{fowler2018}.
    }
\end{figure}

\begin{figure}[h]
    \resizebox{\linewidth}{!}{
        \includegraphics{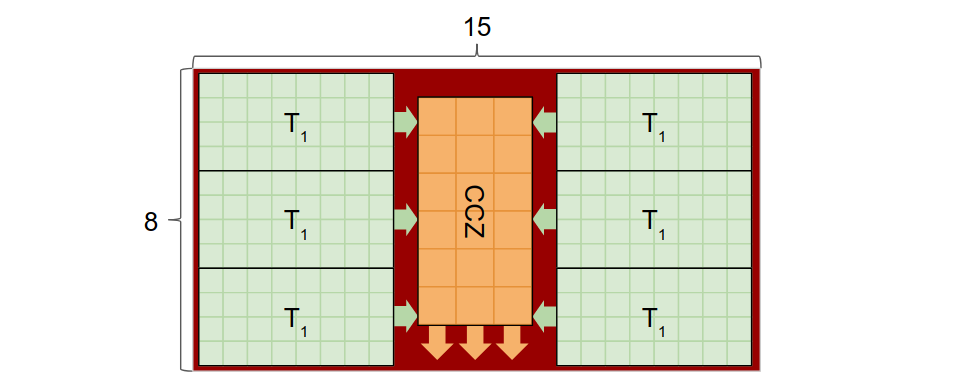}
    }
    \caption{
        \label{fig:ccz-footprint}
        The CCZ factory from \cite{gidney2018magic}, after adjusting the level 1 distance to 17 and the level 2 distance to $d=27$, has a footprint that fits inside a $15 \times 8$ rectangular patch of distance $d=27$ logical qubits.
    }
\end{figure}

\begin{figure}[h!]
\resizebox{0.7\textwidth}{!}{
  \includegraphics{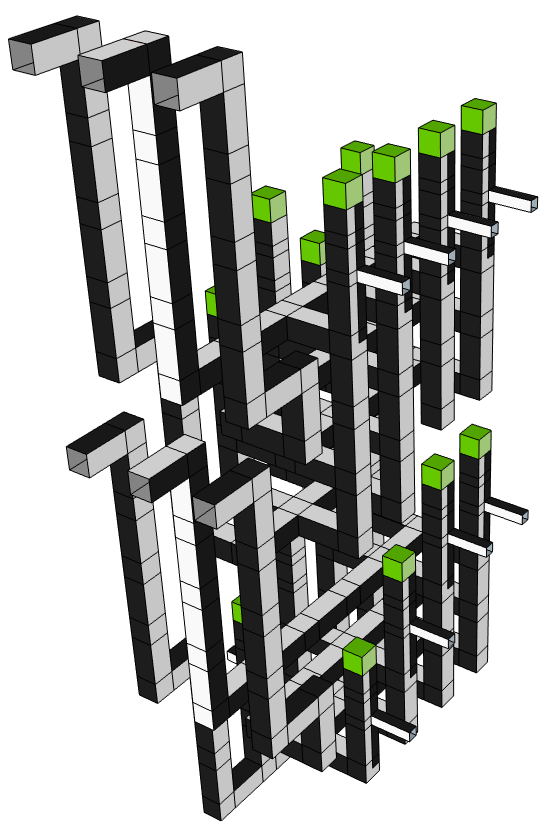}
}
    \caption{
        \label{fig:ccz-factory}
        3D diagram of the core of the CCZ factory factory we are using, derived from the factory in \cite{gidney2018magic}.
        This 3d model is stored in ancillary file ``ccz-factory.skp"; it can be opened online using Sketchup.
        The triplets of open-ended pipes are the CCZ state output.
        Each small open-ended pipe is for a T state input, coming from the six level 1 T factories (not shown here, see \fig{t1-factory}).
        Due to how the factory is interleaved with itself, a CCZ state of sufficient quality is produced every $d \cdot 5$ surface code cycles where $d$ is the code distance.
    }
\end{figure}

\begin{figure}[h]
\resizebox{\textwidth}{!}{
  \includegraphics{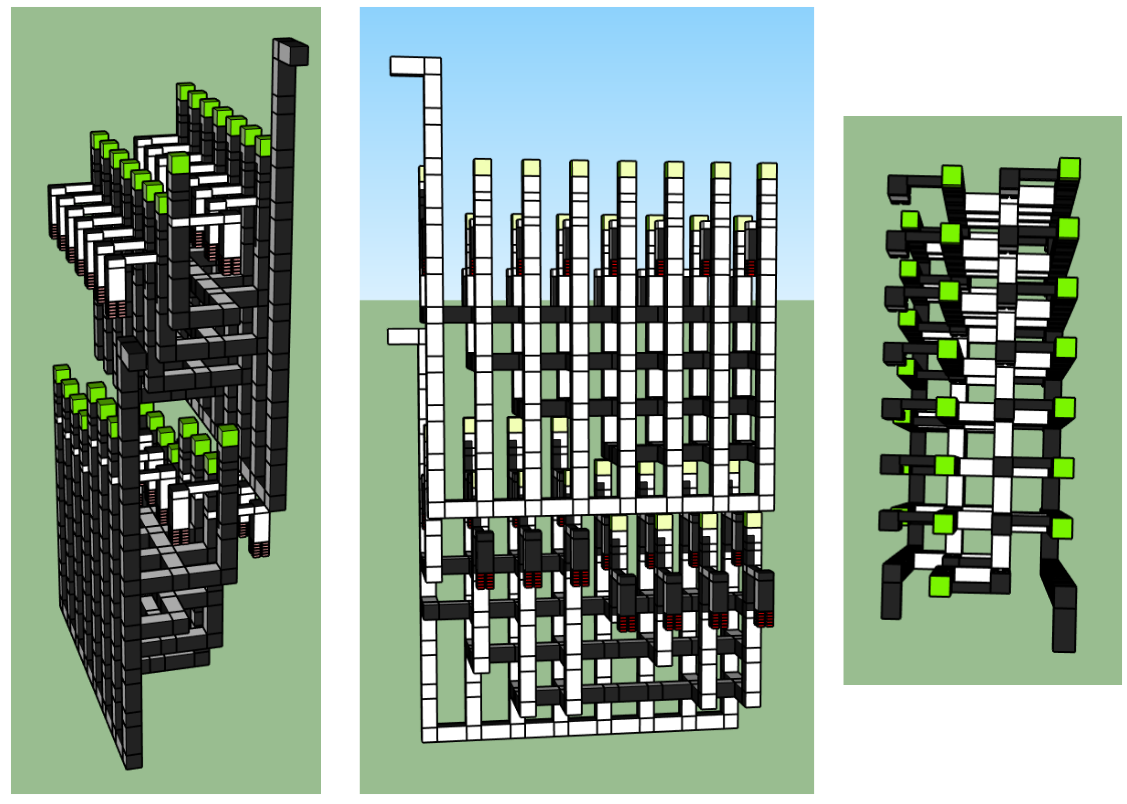}
}
    \caption{
        \label{fig:t1-factory}
        Different views of a 3D diagram of the level 1 T factory we are using to produce $|T_1\rangle$ states, derived from the factory in \cite{fowler2018, gidney2018magic}.
        This 3d model is stored in ancillary file ``t1-factory.skp"; it can be opened online using Sketchup.
        Due to how the factory is interleaved with itself, a $|T_1\rangle$ state is produced every $\distone \cdot 5.75$ surface code cycles (where $\distone$ is the level 1 code distance).
    }
\end{figure}

The exact code distances that we choose depends on the number of operations that have to be performed.
Loosely speaking, the level 2 code distance will almost always be lower (e.g. 27 instead of the original 31) and the level 1 code distance will sometimes be larger (e.g. 17 instead of the original 15).
Changing the code distances requires recomputing the footprint and depth of the factory.
Depending on the ratio between the level 1 and level 2 code distances, the output rate may be limited by either the production rate of level 1 T states or by the height of the level 2 part of the factory.
The footprint of the factory is similarly lower bounded by either one code distance or the other, depending on their relative size (e.g. see \fig{ccz-footprint}).

Since writing \cite{gidney2018magic}, we have checked that it is possible to slightly lower the locations where T states are injected into our factories.
Basically, because the T states are being injected on the outside of the vertical poles storing the factory's qubits, while the stabilizer measurements are happening on the insides of the poles, they are not constrained into waiting for each other and it is possible to overlap the two.
We show this new injection style, which reduces the depth of the CCZ factory from $5.5d$ to $5d$, in \fig{t-inject-overlap}.

We also used this new injection style in the level 1 T factories within the CCZ factory.
Additionally, we found a slightly better way to interleave successive instantiations of the T factories.
In total, as shown in \fig{t1-factory}, we achieved a depth of $5.75d_1$ (compared to $6.25d_1$ in \cite{gidney2018magic}).

For extremely large problem sizes, where the number of Toffoli operations exceeds the capabilities of the CCZ state factory from \cite{gidney2018magic}, we can fall back to the T state factory from \cite{fowler2018}.
According to the spreadsheet included in \cite{gidney2018magic}, and accounting for the ability to increase the code distances, this transition becomes necessary when performing on the order of ten trillion Toffoli operations.

In order to perform a reaction limited computation, more than one factory is needed.
The exact number depends on the reaction time of the control system, the cycle time of the surface code, and the depth of the factory.
In this paper we are assuming a reaction time of 10 microseconds and a cycle time of 1 microsecond.
For the sake of example, we will assume a level 1 code distance of 17 and a level 2 code distance of 27.

The production rate of the factory can be limited by either the level 1 or level 2 distances.
At level 2 the production rate of the factory is limited by the factory's depth times the cycle time times the level 2 code distance.
So the level 2 part of the factory is technically capable of producing states at a rate of $(5 \cdot 1 \mu s \cdot 27)^{-1} \approx 7.4$ kHz.
The level 1 part of the factory needs to produce 8 level 1 T states for each CCZ state that will be output.
There are six level 1 T factories, and they have a depth of $5.75 d_1$, which means the output rate of the entire factory cannot be larger than $(5.75 \cdot 17 \cdot 1 \mu s \cdot 8 / 6)^{-1} \approx 7.7$ kHz.
Therefore the level 2 code distance is the limiting factor, and the factory runs at $7.4$ kHz.

In a reaction limited computation, one CCZ state will be needed per reaction time of the classical control system.
That is to say, CCZ states are consumed at a rate of 100 kHz.
Therefore, given our assumptions, a reaction limited computation requires $\lceil 100/7.4 \rceil = 14$ CCZ factories running in parallel.

Note that the combined footprint of these 14 CCZ factories would be approximately 2.6 million physical qubits.
This suggests that quantum computations covering less than five million physical qubits will see no gains from circuit-level parallelism, such as using carry-lookahead adders instead of ripple-carry adders, because there simply isn't the space to fit the data qubits, the factories, and the routing needed to run multiple threads of execution faster than a single thread of execution.

\section{Reaction limited ripple-carry addition}
\label{sec:adder}

\begin{figure}
\resizebox{\textwidth}{!}{
\Qcircuit @R=1em @C=0.75em {
  &&\lstick{c_\text{in}} &\targ    &\ctrl{1}&\qw      &\qw     &\qw      &\qw     &\qw      &\qw     &\qw     &\qw       &\qw      &\qw       &\qw       &\qw        &\qw       &\qw       &\qw        &\ctrl{1}  &\targ     &\ctrl{1} &\qw &\rstick{c_\text{in}}          &&&&&&\\
  &&\lstick{t_0}         &\targ    &\ctrl{1}&\qw      &\qw     &\qw      &\qw     &\qw      &\qw     &\qw     &\qw       &\qw      &\qw       &\qw       &\qw        &\qw       &\qw       &\qw        &\ctrl{1}  &\qw       &\targ    &\qw &\rstick{(t+i)_0}              &&&&&&\\
  &&\lstick{i_0}         &\ctrl{-2}&\targ   &\targ    &\ctrl{1}&\qw      &\qw     &\qw      &\qw     &\qw     &\qw       &\qw      &\qw       &\qw       &\qw        &\ctrl{1}  &\targ     &\ctrl{1}   &\targ     &\ctrl{-2} &\qw      &\qw &\rstick{i_0}                  &&&&&&\\
  &&\lstick{t_1}         &\qw      &\qw     &\targ    &\ctrl{1}&\qw      &\qw     &\qw      &\qw     &\qw     &\qw       &\qw      &\qw       &\qw       &\qw        &\ctrl{1}  &\qw       &\targ      &\qw       &\qw       &\qw      &\qw &\rstick{(t+i)_1}              &&&&&&\\
  &&\lstick{i_1}         &\qw      &\qw     &\ctrl{-2}&\targ   &\targ    &\ctrl{1}&\qw      &\qw     &\qw     &\qw       &\qw      &\ctrl{1}  &\targ     &\ctrl{1}   &\targ     &\ctrl{-2} &\qw        &\qw       &\qw       &\qw      &\qw &\rstick{i_1}                  &&&&&&\\
  &&\lstick{t_2}         &\qw      &\qw     &\qw      &\qw     &\targ    &\ctrl{1}&\qw      &\qw     &\qw     &\qw       &\qw      &\ctrl{1}  &\qw       &\targ      &\qw       &\qw       &\qw        &\qw       &\qw       &\qw      &\qw &\rstick{(t+i)_2}              &&&&&&\\
  &&\lstick{i_2}         &\qw      &\qw     &\qw      &\qw     &\ctrl{-2}&\targ   &\targ    &\ctrl{1}&\qw     &\targ     &\ctrl{1} &\targ     &\ctrl{-2} &\qw        &\qw       &\qw       &\qw        &\qw       &\qw       &\qw      &\qw &\rstick{i_2}                  &&&&&&\\
  &&\lstick{t_3}         &\qw      &\qw     &\qw      &\qw     &\qw      &\qw     &\targ    &\ctrl{2}&\qw     &\qw       &\targ    &\qw       &\qw       &\qw        &\qw       &\qw       &\qw        &\qw       &\qw       &\qw      &\qw &\rstick{(t+i)_3}              &&&&&&\\
  &&\lstick{i_3}         &\qw      &\qw     &\qw      &\qw     &\qw      &\qw     &\ctrl{-2}&\qw     &\ctrl{1}&\ctrl{-2} &\qw      &\qw       &\qw       &\qw        &\qw       &\qw       &\qw        &\qw       &\qw       &\qw      &\qw &\rstick{i_3}                  &&&&&&\\
  &&\lstick{t_4}           &\qw      &\qw     &\qw      &\qw     &\qw      &\qw     &\qw      &\targ   &\targ   &\qw       &\qw      &\qw       &\qw       &\qw        &\qw       &\qw       &\qw        &\qw       &\qw       &\qw      &\qw &\rstick{(t+i)_4}&&&&&&\\
}
}
    \caption{
        \label{fig:add}
        Cuccaro's ripple-carry adder  \cite{cuccaro2004adder}.
        Uses local operations travelling in a ``V" shaped wave.
        If $m$ is the length of the target register $t$, then the Toffoli count and measurement depth of this construction is $2m-3$.
    }
\end{figure}
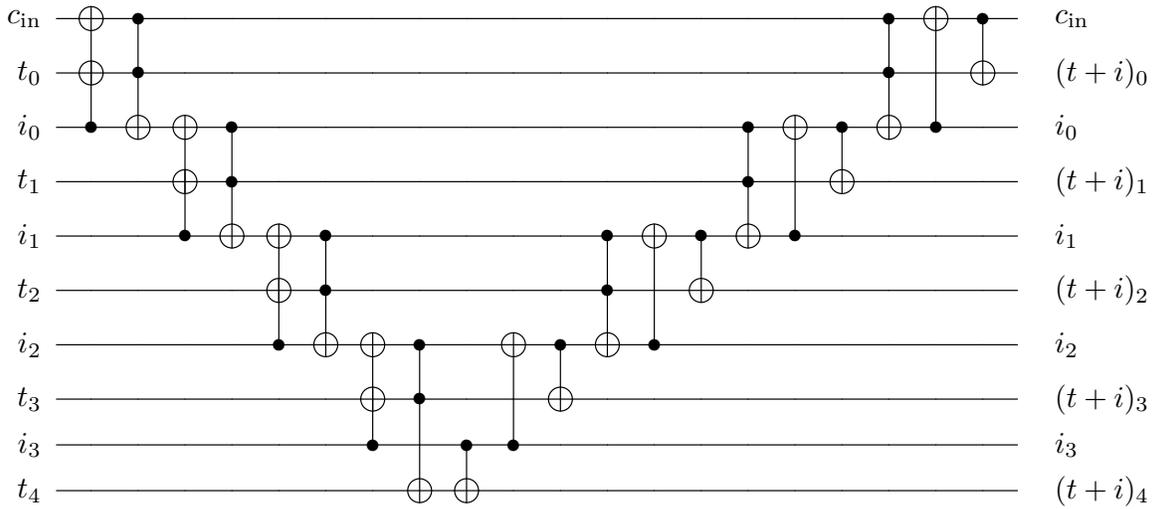

Our goal in this section is to perform a reaction limited addition using the ripple-carry adder described by Cuccaro et al \cite{cuccaro2004adder} (see \fig{add}).
There are two problems we need to solve to make this possible.
First, we need to figure out how to lay out the operations of the adder in a spacelike fashion.
Second, we need to efficiently route the CCZ states being produced by the many CCZ factories into the ripple-carry computation.

\begin{figure}
    \resizebox{\linewidth}{!}{
        \includegraphics{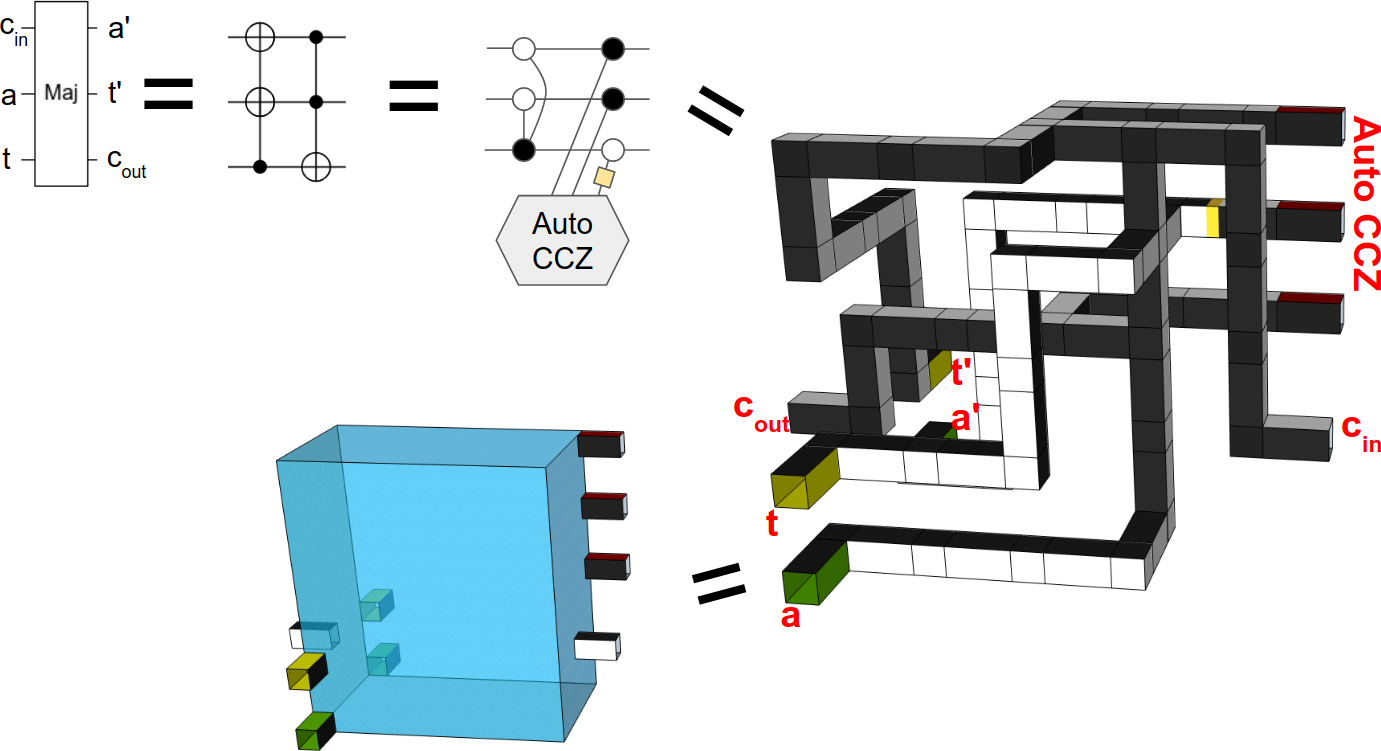}
    }
    \caption{
        \label{fig:maj}
        Lattice surgery implementation of the MAJ operation.
        Top-left is the MAJ circuit from Cuccaro's paper \cite{cuccaro2004adder}.
        Top-center is a ZX calculus graph with function equivalent to the circuit.
        Center-right is a lattice surgery implementation of the ZX graph (stored in ancillary file ``maj.skp" which can be opened online using Sketchup).
        Bottom-center is a simplified representation of the 3d model that shows only the ports and the 3x3x5 bounding box.
    }
\end{figure}

\begin{figure}
    \resizebox{\linewidth}{!}{
        \includegraphics{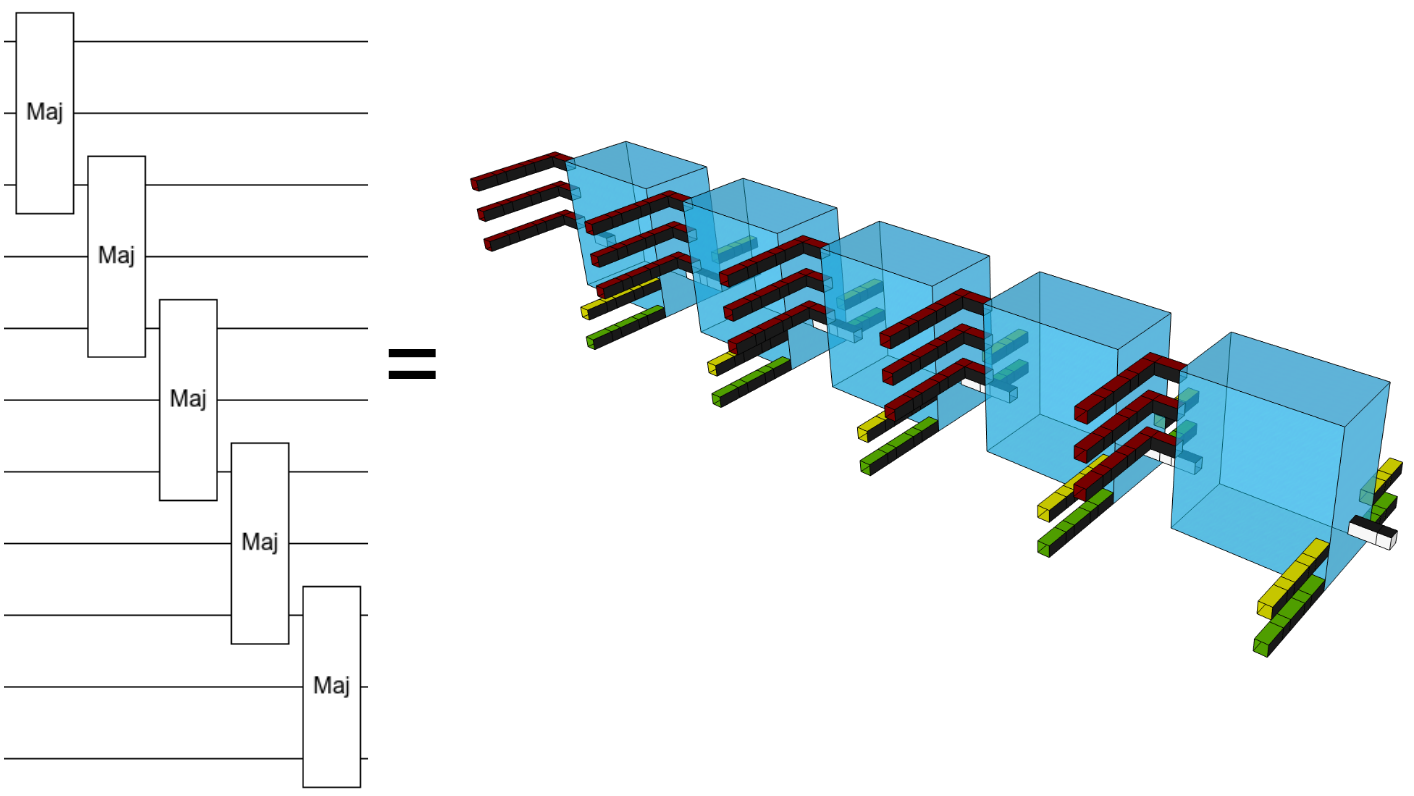}
    }
    \caption{
        \label{fig:maj-sweep}
        Lattice surgery implementation of a series of MAJ operations.
        Each light blue box is an instance of \fig{maj}.
        To enable parallelization, the carry propagation is done left to right, through space, instead of forward through time.
        Dark and light coloring indicates boundary type, but otherwise coloring is for labelling.
        Red bars are injecting three qubits from an AutoCCZ state, green bars are passing qubits from the target register through the block (front to back), and yellow bars are passing qubits from the offset register through the block (front to back).
    }
\end{figure}

Thanks to the flexibility of the AutoCCZ state, laying out the ripple-carry process in a spacelike fashion is straightforward.
We just need implementations of the MAJ and UMA carry rippling operations that accept a CCZ state and propagate the involved bits horizontally across space, instead of vertically through time.
\fig{maj} shows a 3d topological diagram of a MAJ implementation with this property.
It consumes one AutoCCZ state, propagates carry bits left to right, propagates the qubits being operated on front to back, and propagates nothing past-to-future.
It is possible to tile multiple copies of this block across space, as shown in \fig{maj-sweep}.
The control system will then analyze the dependencies between the AutoCCZ states, and propagate the carry signal in a reaction limited fashion (as in \fig{toffoli-sweep}).

Note that we can't lay out the \emph{entire} addition in a spacelike fashion.
That would require a number of factories proportional to the size of the addition, instead of proportional to the reaction time of the control system.
We instead zig-zag the addition back and forth across space, performing just enough carry rippling to keep the factories and the control system busy.

We now move on to the routing question.
Each MAJ box has four inputs and three outputs.
One of the inputs, and also one of the outputs, is the carry qubit.
Another two of the inputs (and outputs) are the data qubits, one from the target register and one from the offset register.
The remaining input is the three qubits making up the CCZ part of an AutoCCZ state.
We must route all these input and output qubits in a way that causes them to intersect the MAJ box at the right place and at the right time.

Our solution to this problem is to zig-zag the carry qubit back and forth along the X axis (right/left through space), while running data qubits through along the Y axis (forward/back through space).
We place CCZ factories in front of and behind the zig-zagging area, so that their outputs are produced directly adjacent to where they are needed making routing trivial.
We then leave small gaps between adjacent factories, so that data qubits from outside the zig-zagging area can be routed through those gaps as needed.

As more and more data qubits are routed from behind the zig-zagging area to in front (or vice versa), we gradually shift the zig-zagging area backward (or forward).
We interleave the two data registers into alternating rows, so that qubits that need to reach the same MAJ box at the same time are adjacent.
Within each row we do additional interleaving, spacing out qubits that are technically sequential in the register.
This prevents traffic jams as the data qubits are routed through the gaps between the factories.

\begin{figure}
    \resizebox{\linewidth}{!}{
        \includegraphics{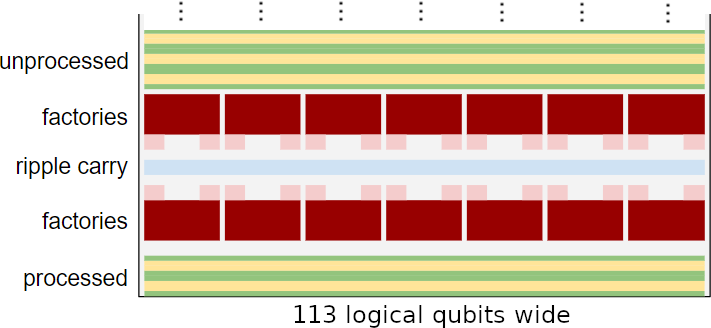}
    }
    \captionof{figure}{
        \label{fig:addition-layout-2d}
        Layout of a reaction limited ripple-carry addition, assuming a level 2 code distance of 27 and a level 1 code distance of 17.
        The CCZ factories (red boxes) and CCZ fixups (pink boxes) are producing AutoCCZ states fed into the MAJ computations (light blue region).
        As the carry qubit zig-zags back and forth, qubits from the offset register and target register (green and yellow rows) are routed through gaps between the factories.
    }
\end{figure}

\begin{figure}[t]
    \resizebox{\linewidth}{!}{
        \includegraphics{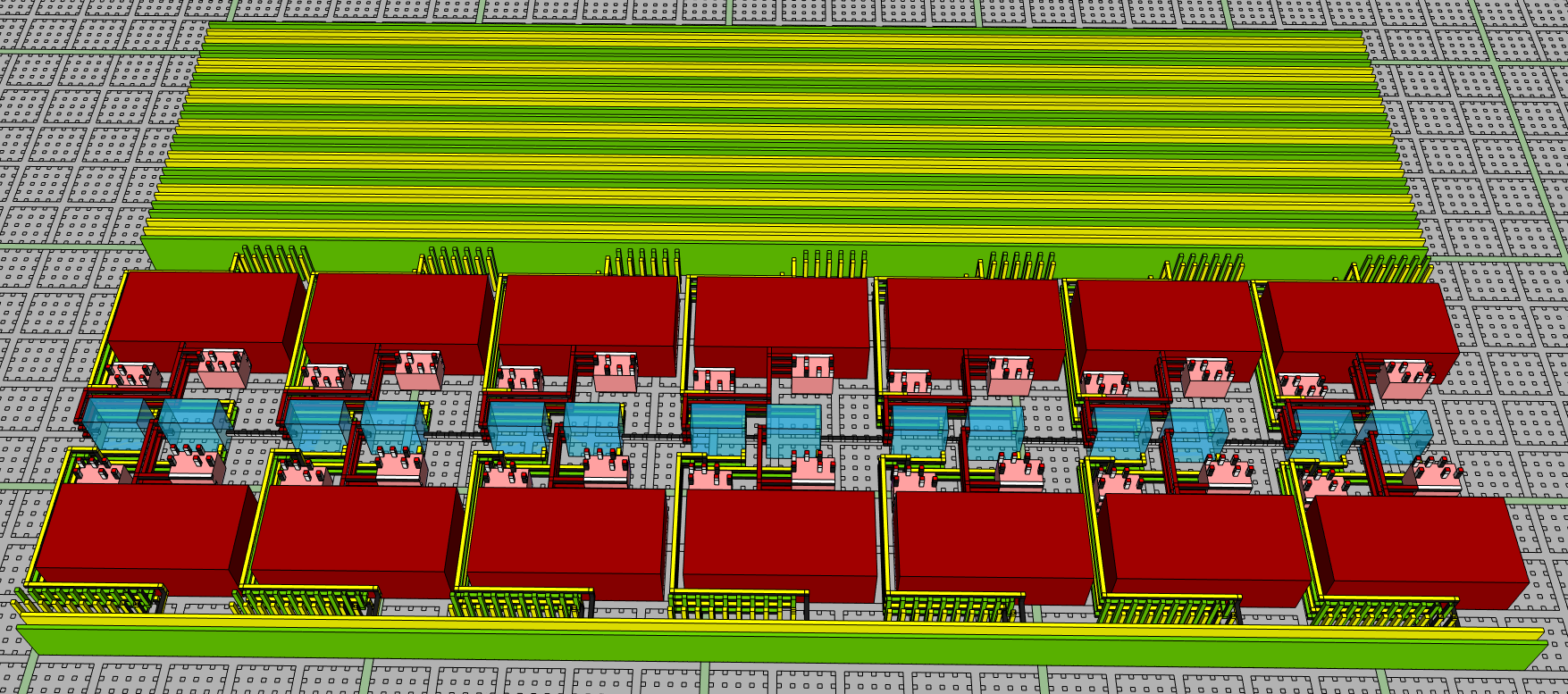}
    }
    \captionof{figure}{
        \label{fig:addition-layout-3d}
        Surface code activity during an addition, assuming a level 2 code distance of 27 and a level 1 code distance of 17.
        This 3d model is stored in ancillary file ``adder-layout.skp", which can be opened online using Sketchup.
        Yellow and green rows are qubits from the target and input registers of the addition.
        Dark blue rows are qubits from an idle register.
        Red boxes are CCZ magic state factories, auto-corrected by the pink boxes (see \fig{delayed-choice-cz}).
        Light blue boxes are the MAJ operation of Cuccaro's adder (see \fig{maj}), arranged into a spacelike sweep to keep ahead of the control software (see \sec{auto-ccz}).
        The full adder is formed by repeating this pattern, with the operating area gradually sweeping up and then down through the green/yellow data (see \fig{time-bars}).
    }
\end{figure}

The layout we are describing can be seen in 2d in \fig{addition-layout-2d} and in 3d in \fig{addition-layout-3d}.
We also provide an ``across time" view in \fig{time-bars}.
Note how there is some additional space for the CCZ fixup boxes from \fig{auto-ccz-fixup-lattice} and for left/right routing of data qubits.
There are two CCZ fixup boxes per CCZ factory because the routing qubits (the ``chimneys") emerging from a CCZ fixup box can extend vertically into the next ``zig-zag layer" before the control system determines how to measure said qubits.

This layout is capable of performing ripple-carry additions at the reaction limited rate, propagating the carry information from qubit to qubit at 100kHz.
Given our physical assumptions, this layout would add a pair of thousand-qubit registers in approximately 20 milliseconds.

\section{Clifford limited table lookups}
\label{sec:lookup}

A ripple-carry adder is ideal for reaction limited computation because it has only a small amount of Clifford operations per Toffoli operation.
A table lookup (also called a QROM read \cite{babbush2018}) is exactly the opposite.
For each Toffoli performed in a table lookup, there is a huge amount of Clifford work to do.
In particular, each Toffoli triggers a gigantic multi-target CNOT potentially touching all of the lookup's output qubits.
Because of this, the limiting factor during a table lookup is not the control system's reaction time but rather access to the output qubits.

In order to target a logical qubit with a CNOT, there must be an unused logical-qubit sized patch of surface code adjacent to that logical qubit.
The CNOT operation will then need to use that patch for $d$ cycles, where $d$ is the code distance.
For qubits where only one side is accessible, only one CNOT can be performed per $d$ cycles.
Given our assumption of a surface code cycle time of 1 microsecond, and using a code distance of 27 just for example, this suggests a maximum CNOT rate of 37kHz (instead of the 100kHz of a reaction limited computation).

It is often possible to work around this CNOT rate limitation.
For example, if there are multiple single-control single-target CNOTs all targeting the same qubit, it is possible to fuse the many CNOTs into a single generalized CNOT where the control is a Pauli product of all the individual controls.
But this doesn't work in the case of table lookups, because the set of relevant control qubits differs from output qubit to output qubit.

One thing we can do is to make two sides of each qubit accessible, instead of just one.
The gigantic multi-target CNOTs can then alternate between using one side, and using the other side.
This doubles the achievable Toffoli rate from 37kHz to 74kHz, which is much closer to 100kHz.
One can go further, e.g. making temporary copies of qubits in order to allow more and more parallel access, but the space tradeoffs start to become problematic.
In this paper we will limit ourselves to two-sided access.

We focus on the case where we are performing a large lookup that prepares a register whose rows are interleaved between rows of another register.
This specific case is interesting to us because it is the kind of lookup that is needed when performing windowed arithmetic \cite{gidney2019windowedarithmetic}, where groups of addition operations are replaced by single lookup+addition operations.

While performing the lookup, we use a tiled row interleaving pattern of $R\_L\_L\_R$ where an $L$ is a lookup data row, an $R$ is an existing data row not involved in the lookup, and an underscore is a empty access row that can be used when performing the multi-target CNOTs.
The multi-target CNOTs alternate between using the single inner access row and both of the outer access rows.
In order to access the access rows, we place cross-row access corridors on opposing sides of the layout.
The multi-target CNOTs alternate between using the two access corridors, so that they can branch into individual access rows as needed.

\begin{figure}
    \resizebox{\linewidth}{!}{
        \includegraphics{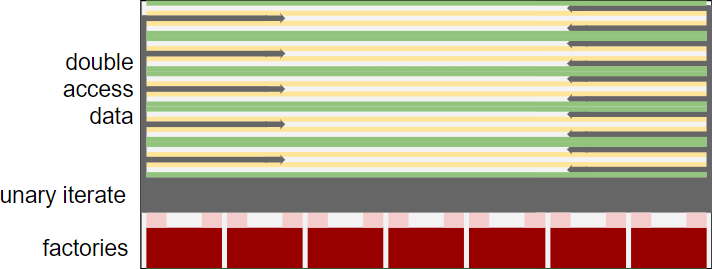}
    }
    \captionof{figure}{
        \label{fig:lookup-layout}
        Data layout during a double-access lookup, assuming a level 2 code distance of 27 and a level 1 code distance of 17.
        Output qubits (yellow) are interleaved between the rows of an existing register (green).
        The vertical access corridors and horizontal access rows provide two distinct ways to simultaneously access all output qubits when performing many-target CNOTs.
        The dark gray area in the bottom region is sufficient space to run a unary iteration \cite{babbush2018}.
        The number of factories shown is slightly lower than what is needed to saturate the access rate of the double-access hallways.
    }
\end{figure}

\begin{figure}[ht]
    \resizebox{\linewidth}{!}{
        \includegraphics{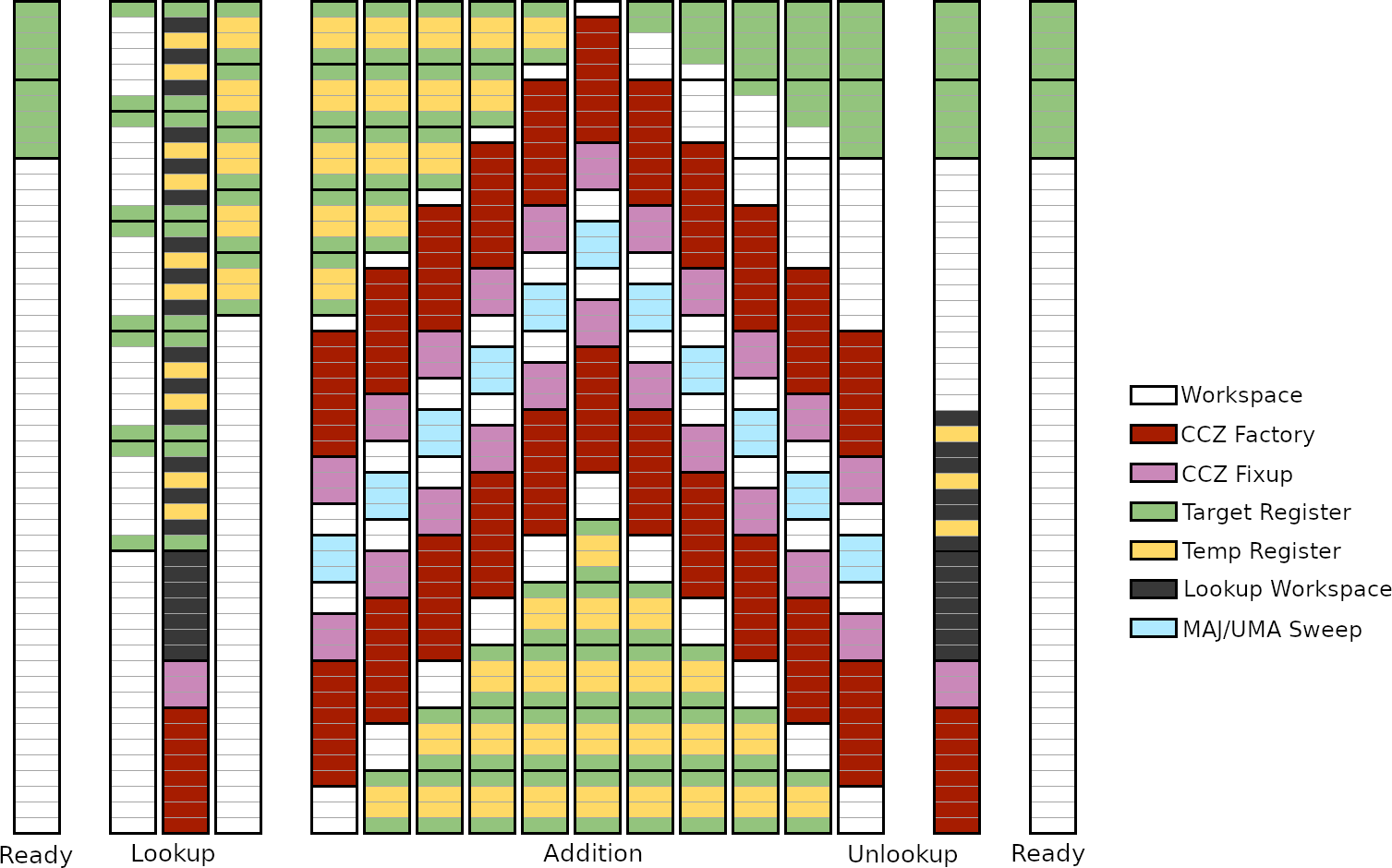}
    }
    \caption{
        \label{fig:time-bars}
        Layout of data (top to bottom) over time (left to right) during a lookup addition.
        During lookup, the target register is spread out to make room for the temporary register that will hold the lookup's output.
        During addition the target register and lookup output register are squeezed through a moving operating area that sweeps up then down, applying the MAJ and UMA sweeps of Cuccaro's adder.
        Uncomputing the lookup is done using measurement based uncomputation \cite{berry2019qubitization}, and has negligible cost compared to the other steps.
    }
\end{figure}

\fig{lookup-layout} has a 2d floor plan of where factories and access hallways are located during a lookup.
We also provide an ``across time" view in \fig{time-bars} of a lookup followed by an addition.

\section{Conclusion}
\label{sec:conclusion}

In this paper we defined a self-correcting AutoCCZ state, improved the distillation factory and delayed choice CZ constructions underlying it, and demonstrated the basics of using it to lay out low-depth surface code computations.
We laid out a reaction limited ripple-carry adder, a Clifford limited table lookup, and ensured they fit together in a fashion that would be useful for performing windowed arithmetic.

We estimated, under plausible physical assumptions for future large-scale superconducting qubit platforms, that running a serial circuit at the reaction limited rate requires 14 CCZ factories.
Because this number of factories covers several million physical qubits, we believe that (under our physical assumptions) computations involving fewer than five million physical qubits will not benefit from using circuits that perform Toffolis in parallel (as in carry-lookahead adders) instead of serially (as in ripple-carry adders).

We hope readers find the techniques we've described, and the example layouts we've presented, helpful when laying out their own large scale computations.

\section{Contributions}

Craig Gidney constructed the AutoCCZ state and produced the adder and lookup layouts using it.
Austin Fowler helped iterate on ideas for laying out CCZ fixups, found the more efficient factory interleavings, and played a supervisory role guiding the content of the paper.

\section{Acknowledgements}

We thank Hartmut Neven for creating an environment where this research was possible in the first place.

\bibliographystyle{plainnat}
\bibliography{refs}

\begin{thebibliography}{15}
\providecommand{\natexlab}[1]{#1}
\providecommand{\url}[1]{\texttt{#1}}
\expandafter\ifx\csname urlstyle\endcsname\relax
  \providecommand{\doi}[1]{doi: #1}\else
  \providecommand{\doi}{doi: \begingroup \urlstyle{rm}\Url}\fi

\bibitem[Babbush et~al.(2018)Babbush, Gidney, Berry, Wiebe, McClean, Paler,
  Fowler, and Neven]{babbush2018}
Ryan Babbush, Craig Gidney, Dominic~W Berry, Nathan Wiebe, Jarrod McClean,
  Alexandru Paler, Austin Fowler, and Hartmut Neven.
\newblock Encoding electronic spectra in quantum circuits with linear t
  complexity.
\newblock \emph{Physical Review X}, 8\penalty0 (4):\penalty0 041015, 2018.

\bibitem[Berry et~al.(2019)Berry, Gidney, Motta, McClean, and
  Babbush]{berry2019qubitization}
Dominic~W Berry, Craig Gidney, Mario Motta, Jarrod~R McClean, and Ryan Babbush.
\newblock Qubitization of arbitrary basis quantum chemistry by low rank
  factorization.
\newblock \emph{arXiv preprint arXiv:1902.02134}, 2019.

\bibitem[Bravyi and Kitaev(2005)]{bravyi2005distillation}
Sergey Bravyi and Alexei Kitaev.
\newblock Universal quantum computation with ideal clifford gates and noisy
  ancillas.
\newblock \emph{Physical Review A}, 71\penalty0 (2):\penalty0 022316, 2005.

\bibitem[Cuccaro et~al.(2004)Cuccaro, Draper, Kutin, and
  Moulton]{cuccaro2004adder}
Steven~A Cuccaro, Thomas~G Draper, Samuel~A Kutin, and David~Petrie Moulton.
\newblock A new quantum ripple-carry addition circuit.
\newblock \emph{arXiv preprint quant-ph/0410184}, 2004.

\bibitem[de~Beaudrap and Horsman(2017)]{de2017zxlattice}
Niel de~Beaudrap and Dominic Horsman.
\newblock The zx calculus is a language for surface code lattice surgery.
\newblock \emph{arXiv preprint arXiv:1704.08670}, 2017.

\bibitem[Eastin(2013)]{eastin2013distilling}
Bryan Eastin.
\newblock Distilling one-qubit magic states into toffoli states.
\newblock \emph{Physical Review A}, 87\penalty0 (3):\penalty0 032321, 2013.

\bibitem[Fowler et~al.(2012)Fowler, Mariantoni, Martinis, and
  Cleland]{fowler2012surfacecodereview}
A.~G. Fowler, M.~Mariantoni, J.~M. Martinis, and A.~N. Cleland.
\newblock Surface codes: Towards practical large-scale quantum computation.
\newblock \emph{Phys. Rev. A}, 86:\penalty0 032324, 2012.
\newblock URL \url{https://doi.org/10.1103/PhysRevA.86.032324}.
\newblock arXiv:1208.0928.

\bibitem[Fowler(2012)]{fowler2012time}
Austin~G Fowler.
\newblock Time-optimal quantum computation.
\newblock \emph{arXiv preprint arXiv:1210.4626}, 2012.

\bibitem[Fowler and Gidney(2018)]{fowler2018}
Austin~G Fowler and Craig Gidney.
\newblock Low overhead quantum computation using lattice surgery.
\newblock \emph{arXiv preprint arXiv:1808.06709}, 2018.

\bibitem[Gidney(2019)]{gidney2019windowedarithmetic}
Craig Gidney.
\newblock Windowed quantum arithmetic.
\newblock \emph{arXiv preprint arXiv:1905.07682}, 2019.

\bibitem[Gidney and Fowler(2018)]{gidney2018magic}
Craig Gidney and Austin~G Fowler.
\newblock Efficient magic state factories with a catalyzed ccz to 2t
  transformation.
\newblock \emph{arXiv preprint arXiv:1812.01238}, 2018.

\bibitem[Gottesman and Chuang(1999)]{gottesman1999gateteleport}
Daniel Gottesman and Isaac~L Chuang.
\newblock Demonstrating the viability of universal quantum computation using
  teleportation and single-qubit operations.
\newblock \emph{Nature}, 402\penalty0 (6760):\penalty0 390, 1999.

\bibitem[Jones(2013)]{jones2013}
Cody Jones.
\newblock Low-overhead constructions for the fault-tolerant toffoli gate.
\newblock \emph{Physical Review A}, 87\penalty0 (2):\penalty0 022328, 2013.

\bibitem[Li(2015)]{li2015}
Ying Li.
\newblock A magic state’s fidelity can be superior to the operations that
  created it.
\newblock \emph{New Journal of Physics}, 17\penalty0 (2):\penalty0 023037,
  2015.

\bibitem[Litinski(2018)]{litinski2018}
Daniel Litinski.
\newblock A game of surface codes: Large-scale quantum computing with lattice
  surgery.
\newblock \emph{arXiv preprint arXiv:1808.02892}, 2018.

\end{thebibliography}

\end{document}